# Enhanced Security and Efficiency in Blockchain with Aggregated Zero-Knowledge Proof Mechanisms

**Oleksandr Kuznetsov[1,2], Member, IEEE, Alex Rusnak[1], Anton Yezhov[1], Dzianis Kanonik[1], Kateryna Kuznetsova[1], Stanislav Karashchuk[1]**

[1]Proxima Labs, 1501 Larkin Street, suite 300, San Francisco, USA
[2]Department of Political Sciences, Communication and International Relations, University of Macerata, Via Crescimbeni, 30/32, 62100 Macerata, Italy

Corresponding author: Oleksandr Kuznetsov (e-mail: kuznetsov@karazin.ua; ORCID: 0000-0003-2331-6326).

This project is supported by Proxima Labs, 1501 Larkin Street, suite 300, San Francisco, USA

**ABSTRACT** Blockchain technology has emerged as a revolutionary tool in ensuring data integrity and security in digital transactions. However, the current approaches to data verification in blockchain systems, particularly in Ethereum, face challenges in terms of efficiency and computational overhead. The traditional use of Merkle Trees and cryptographic hash functions, while effective, leads to significant resource consumption, especially for large datasets. This highlights a gap in existing research: the need for more efficient methods of data verification in blockchain networks. Our study addresses this gap by proposing an innovative aggregation scheme for Zero-Knowledge Proofs within the structure of Merkle Trees. We develop a system that significantly reduces the size of the proof and the computational resources needed for its generation and verification. Our approach represents a paradigm shift in blockchain data verification, balancing security with efficiency. We conducted extensive experimental evaluations using real Ethereum block data to validate the effectiveness of our proposed scheme. The results demonstrate a drastic reduction in proof size and computational requirements compared to traditional methods, making the verification process more efficient and economically viable. Our contribution fills a critical research void, offering a scalable and secure solution for blockchain data verification. The implications of our work are far-reaching, enhancing the overall performance and adaptability of blockchain technology in various applications, from financial transactions to supply chain management.

## I. INTRODUCTION

The domain of blockchain technology has rapidly emerged as a pivotal force in modern digital ecosystems, revolutionizing how data is stored, transactions are processed, and trust is established in decentralized networks [1]. As the cornerstone of cryptocurrencies like Bitcoin and platforms like Ethereum, blockchain technology has expanded far beyond its initial financial applications to sectors including healthcare, supply chain management, and digital identity verification [2], [3].

Despite its widespread adoption, the blockchain field confronts significant challenges, particularly in verifying the integrity and authenticity of the vast volumes of data these networks process [4], [5]. Traditional mechanisms, such as the use of Merkle Trees and cryptographic hashes, though fundamental, reveal limitations in handling the increased scale and complexity of modern blockchain networks [6], [7]. These challenges are more pronounced in versatile and expansive platforms like Ethereum, which support complex smart contracts and decentralized applications (dApps) [8], [9].

Current research extensively explores various blockchain dimensions but often falls short in effectively addressing the scalability and efficiency of data verification processes [6]. As blockchain networks grow in size and complexity, the computational and storage requirements for traditional verification methods, like complete Merkle Trees and hash calculations, become increasingly impractical. This gap in research points to the need for innovative solutions capable of scaling with the network while ensuring data integrity and transactional security.

Our research is centered around the critical issue of enhancing the efficiency and scalability of data verification within blockchain networks. This challenge is particularly vital in the context of Ethereum, a platform known for its wide array of functionalities, including complex smart contracts and a robust dApp ecosystem [10]. As Ethereum continues to evolve, the volume of transactions and the complexity of interactions within its blockchain increase exponentially [11], [12].

The primary problem lies in the traditional verification methods, which, while robust, are not optimized for large-scale applications [7], [13]. These methods can lead to bottlenecks, increased transaction costs, and delays, ultimately hampering the network's performance and user experience. For instance, the verification of transactions and smart contracts in Ethereum currently requires significant computational resources and time, especially when dealing with complete Merkle Trees for every block and transaction [14], [15]. This issue is further compounded by the network's growing size and the constant addition of new blocks and transactions.

Addressing this problem is not merely a technical challenge but also a critical requirement for the broader adoption and sustainability of blockchain technology. As Ethereum and similar platforms expand their scope and scale, the need for more efficient, cost-effective, and scalable verification methods becomes increasingly imperative [11]. Our research aims to fill this gap by proposing an innovative approach that not only enhances the efficiency of the verification process but also maintains, if not strengthens, the security and integrity of the blockchain.

The central theme of this research is the application of Zero-Knowledge Proofs (ZKPs) in the realm of blockchain technologies [16], [17], specifically as an alternative to Merkle proofs in the Ethereum network. The emphasis is on the capabilities of ZKPs to reduce the size of proofs and accelerate their verification process.

Merkle proofs play a pivotal role in maintaining the integrity and security of blockchains, providing a means for efficient verification of large data sets without necessitating their complete transfer [18], [19]. However, this mechanism has its limitations, particularly in the context of the increasing data volume and processing speed requirements in modern blockchain systems. ZKPs present a promising solution to these challenges, offering data verification without revealing the information itself, thereby significantly enhancing the efficiency and scalability of the system [20], [21].

### A. OUR CONTRIBUTION TO THE FIELD
We propose a novel ZKP scheme that leverages advanced cryptographic techniques to ensure data integrity and transaction authenticity in blockchain systems. Our approach addresses the scalability and efficiency issues by reducing the size and computational overhead of the proofs, thereby facilitating faster and cost-effective verification. This contribution is a significant step towards enhancing the practicality and scalability of blockchain verification processes, especially in complex networks like Ethereum.

### B. STRUCTURE OF THE ARTICLE
This article is structured to provide a comprehensive analysis and presentation of our research findings. Following the introduction, we delve into the state of the art in blockchain technology, focusing on the key developments and existing limitations within this field. We then transition to the background section, which lays the foundational concepts crucial to understanding our research. This part covers cryptographic hash functions, Merkle Trees, and ZKPs, explaining their relevance and application in the blockchain context.

The core of the article is dedicated to our methods and proposed solutions. We start by exploring the intricacies of the Merkle tree structure and its application in blockchain data verification. This discussion leads to the introduction of our novel aggregation scheme for ZKPs and the concept of recursive zk-SNARK proofs. We follow this with a detailed presentation of our proposed ZKP Scheme, highlighting its potential to revolutionize data verification processes in blockchain networks.

To validate our theoretical propositions, we present an experimental evaluation of the proposed zk-Proof System. This section is meticulously structured to cover our research methodology, the results from zk-STARK and zk-SNARK experiments, and a thorough security analysis of the proposed aggregation scheme. Each experiment is contextualized within the real-world scenario of the Ethereum blockchain, ensuring practical relevance and applicability.

The discussion of research findings offers a critical analysis of our results, weaving together the theoretical and experimental aspects to present a cohesive understanding of the implications of our work. The article concludes with a summary of our research, its significance in the blockchain field, and potential avenues for future work.

Lastly, we include an appendix that details the mathematical underpinnings of calculating collision probability in hash functions, providing additional depth and clarity to our research methodology.

## II. STATE OF THE ART
Yang and Li (2020) [16] investigated a blockchain-based digital identity management scheme using ZKPs. Their scheme aids in protecting identity attributes, but our research is directed towards a more extensive application of ZKPs in blockchain, including optimizing data processing and verification.

Huang et al. (2022) [17] developed a blockchain-based continuous data integrity checking protocol with zero-knowledge privacy protection. While their work concentrates on protecting data in cloud storage, our research extends the application of such principles to enhance data processing efficiency in blockchains.

Di Francesco Maesa et al. (2023) [22] explored self-sovereign and blockchain-based access control systems supporting attribute privacy using ZKPs. While this research advances the idea of transparent policy evaluation without disclosing sensitive attributes, our work focuses on extending the application of ZKPs to improve scalability and efficiency in more extensive blockchain applications.

Emami et al. (2023) [23] proposed a scalable decentralized e-voting system utilizing zero-knowledge off-chain computations. Their approach fosters transparency and privacy; however, our research seeks to expand these concepts across a broader spectrum of blockchain applications, including transactions and data processing.

Rodinko, Oliynykov, and Nastenko (2023) [24] introduced a decentralized Proof-of-Burn auction for secure cryptocurrency upgrades. Although their approach is unique in the cryptocurrency domain, our work aims to employ similar concepts for improving data verification in blockchain, broadening the application of such mechanisms.

The following two works are the closest to our research.

Loporchio, Bernasconi, Di Francesco Maesa, and Ricci (2023) [25] provided a comprehensive survey of set accumulators in blockchain systems. Set accumulators are cryptographic primitives that represent large sets of elements with a single constant-size value and efficiently verify the membership of a value in that set. The authors discussed various accumulator constructions from a complexity perspective and surveyed their applications in blockchain technology, particularly focusing on query authentication, stateless transaction validation, anonymity enhancement, and identity management. While this review underscores the importance of accumulators in enhancing anonymity and identity management in blockchain, our research aims to complement these approaches by delving deeper into the application of ZKPs for more comprehensive data protection and verification process optimization.

Boo, Kim, and Ko (2022) [26] introduced LiteZKP, a framework for supporting multiple anonymous payments using a smart contract-based ZKP protocol on resource-limited devices. LiteZKP includes novel schemes, such as a new Merkle tree mechanism, to reduce the burden of ZKP operations and integrates smart contract-based ZKP with an off-chain payment channel to minimize the amount of ZKP operations in continuous data exchanges. While this approach reduces latency and energy consumption on IoT/mobile edge computing platforms, our research is directed toward a broader application of ZKPs within the blockchain context, including data verification and processing optimization.

Analyzing these studies reveals a concentration on specific aspects of security, anonymity, and efficiency in various blockchain applications. However, there is a gap in the comprehensive application of ZKPs for optimizing verification and data processing across a wide range of blockchain applications. Our research aims to fill this gap by proposing a new mechanism based on ZKPs that offers enhanced security, privacy, and scalability in blockchain systems. We aspire to extend the boundaries of ZKP application, demonstrating their efficacy not only in specific cases but also in a more general context of blockchain technologies. This endeavor positions our research at the forefront of exploring ZKPs in a broad spectrum of blockchain applications, pushing the limits of their utility and effectiveness.

## III. BACKGROUND

The evolution of blockchain technology is inextricably linked to the improvement of mechanisms for data security and verification. In this context, Merkle proofs, a data structure enabling efficient and secure verification of large data arrays, assume a pivotal role. This method, developed by Ralph Merkle in 1980 [27], [28], has found widespread application in various cryptographic systems, including blockchain and cryptocurrencies.

However, as the volume of data and the complexity of transactions in blockchain systems increase, particularly in extensive networks like Ethereum, the limitations of Merkle proofs become increasingly apparent [29], [30]. The issue lies in the necessity to store and process a separate Merkle path for each transaction being verified, leading to significant consumption of computational and storage resources. This becomes critically problematic in scenarios requiring data verification over extended periods or when numerous transactions occur within short intervals.

This study examines an alternative approach using ZKPs, which offers a solution to the aforementioned challenges. ZKPs, a form of cryptographic verification, enable the proving of the existence or truth of information without revealing the information itself [31], [32]. This approach promises a substantial reduction in the volume of data required for verification, consequently enhancing the speed and efficiency of transaction processing in the blockchain.

### A. CRYPTOGRAPHIC HASH FUNCTIONS

Cryptographic hashing plays a pivotal role in information security, including blockchain and cryptocurrency applications [33]. A cryptographic hash function $H$ transforms input data $x$ of arbitrary length into a fixed-length output $H(x)$, known as the hash value. Key properties of cryptographic hash functions include:

- Determinism. For any input $x$, the hash function consistently produces the same hash value $H(x)$.
- High Computational Efficiency. For any given input $x$, computing $H(x)$ should be fast.
- Avalanche Effect. Minor changes in input data $x$ should lead to substantial and unpredictable changes in the hash value $H(x)$.

- One-way Function. For a given hash value $y$, it should be computationally infeasible to find an input $x$ such that $H(x) = y$.
- Collision Resistance. It should be practically impossible to find two different inputs $x_1$ and $x_2$ such that $H(x_1) = H(x_2)$.

These characteristics make cryptographic hash functions critically important for ensuring data integrity and security in blockchain systems. They are used for creating digital signatures, verifying data integrity, and in proof-of-work mechanisms in the context of cryptocurrencies.

### B. MERKLE TREES

Merkle trees, also known as hash trees, play a crucial role in cryptography and blockchain technologies. They provide an efficient and secure method for verifying large amounts of data. The structure of a Merkle tree is based on the principles of hashing and forms a binary tree, where each leaf contains the hash of an individual data block, and each internal node contains the hash derived from the combination of hashes of its child nodes.

The formalization of Merkle trees can be represented by the following elements [27], [28]:

- Leaf Nodes: For data blocks $d_1, d_2, \ldots, d_n$, the leaf nodes are defined as

$$H(d_1), H(d_2), \ldots, H(d_n),$$

where $H$ is a cryptographic hash function.

- Internal Nodes: Each internal node $v$ is the hash of the combination of its child nodes. For example, if a node $v$ has child nodes $a$ and $b$, then $v = H(a \| b)$, where $\|$ denotes concatenation.
- Root of the Tree: The root of the Merkle tree $r$ is a hash that aggregates information from all leaf nodes. $r$ provides a compact representation of the entire data set.

Key Properties of Merkle Trees:

- Data Verification: To confirm the inclusion of data $d_i$ in the set, it suffices to verify the correspondence of the path from $H(d_i)$ to the root $r$.
- Storage Optimization: Storing only the root $r$ allows for data verification while minimizing the need to store the entire data set.
- Security: Any change in data $d_i$ results in a change in the corresponding hash at the leaf node and, consequently, throughout the path to the root $r$, facilitating the detection of modifications.
- Scalability: Adding or changing data $d_i$ requires recalculating only specific paths in the tree, not the entire tree, making Merkle trees efficient for processing large data sets.

### C. ZERO-KNOWLEDGE PROOFS

ZKPs represent a pivotal concept in modern cryptography and blockchain technology, offering robust tools for enhancing privacy and security. These are specialized cryptographic methods that allow a prover to convince a verifier of the truth of a certain statement without revealing any additional information.

Key Properties of ZKPs [20]:

1. Zero-Knowledge: If a statement $S$ is true, the verifier only learns the fact of its truthfulness, without gaining any other information. Mathematically, this can be expressed as: if $S$ is true, then $P(S) \rightarrow V = True$, where $P$ is the prover and $V$ is the verifier.
2. Completeness: If a statement $S$ is true, an honest prover can always convince an honest verifier of its truth. That is, $\forall S : S = True \rightarrow P(S) \rightarrow V = True$.
3. Soundness: If a statement $S$ is false, a dishonest prover cannot convince an honest verifier of its truth, except with a negligible probability of error. This implies $\forall S : S = False \rightarrow P(S) \rightarrow V = False$ with high probability.

In the context of blockchain and cryptocurrencies, ZKPs are used to ensure transaction anonymity and data confidentiality [34]. Examples include protocols such as zk-SNARKs (Zero-Knowledge Succinct Non-Interactive Argument of Knowledge) and zk-STARKs (Zero-Knowledge Scalable Transparent Argument of Knowledge), which provide high levels of privacy and security for transactions and data verification.

Next, we will discuss how hashing, Merkle trees, and ZKPs are integrated into our proposed new data verification mechanism in blockchain systems, enhancing security and efficiency.

## IV. METHODS

In this section, we delve into the key methodologies and concepts that form the foundation of our research. We thoroughly examine technologies such as Merkle Trees, ZKPs, aggregation, and recursive proofs, which are fundamental to the development of our system. These methodologies provide a robust and efficient basis for creating an advanced data verification system, particularly for confirming the inclusion of data in blockchain blocks, such as those of Ethereum. Our research aims to develop a system that not only ensures a high level of security and confidentiality but also optimizes computational resources and simplifies the verification process.

### A. MERKLE TREE

Merkle Trees are fundamental data structures used to ensure the integrity and identifiability of data stored in blockchain blocks, such as Ethereum. These mechanisms provide a simple and efficient way to verify the presence of an element in a data set using cryptographic hashing functions.

Merkle Tree (see Fig. 1) is a tree-like structure where each leaf represents a hash of data, and each internal node is the hash of its child nodes. The root of the tree, known as the Merkle Root, is a unique identifier for the entire tree, encompassing information about all the data in the tree.

The process of creating a Merkle Tree (Fig. 1) starts with calculating the hash for each data element. These hashes become the leaves of the tree. The leaves are then paired, and a hash is calculated for each pair, becoming the nodes of the next level. This process is repeated until a single hash remains, forming the root of the Merkle Tree.

Mathematically, this can be represented as follows: for data $d_1, d_2, ..., d_n$, hashes $h_1, h_2, ..., h_n$ are computed as $h_i = H(d_i)$, where $H$ is the hashing function. Then, the nodes of the next level are computed as $h_{ij} = H(h_i \| h_j)$, where $\|$ denotes concatenation.

### B. MERKLE PROOF

Merkle Proof is a mechanism that allows verifying the presence of a specific data element in a Merkle Tree without requiring access to the entire tree (see Fig. 2). This is achieved by providing the path from the leaf to the root along with the corresponding hashes of the sibling nodes.

To generate a Merkle Proof for an element $d_i$, a set of hashes $h = \{h_1, h_2, ..., h_j, ..., h_k\}$ is required, such that starting from $h_i = H(d_i)$ (the hash of $d_i$), the root hash $r$ can be computed by applying the hash function in a specific sequence. This sequence depends on the position of $d_i$ in the data set.

Thus, if there is a Merkle Proof $h = \{h_1, h_2, ..., h_j, ..., h_k\}$ for $d_i$, the root hash can be calculated as

$$r = H(H(...H(H(H(d_i) \| h_1)...) \| h_{k-1}) \| h_k), \quad (1)$$

where $H$ is the hash function. If the computed $r$ matches the root hash of the tree, it confirms that $d_i$ is part of the data set.

This process allows for the local verification of the membership of an element in a large data set, requiring only a small set of additional information and the root hash. For instance, for a data set $D = \{d_1, d_2, ..., d_{2^{16}}\}$, only 16 hashes in the set $h = \{h_1, h_2, ..., h_{2^{16}}\}$ are needed to prove the membership of $d_i$ in $D$.

Analyzing the concept of Merkle Trees and Merkle Proofs, it is essential to highlight the following apparent advantages:
1. Efficient Data Verification: Merkle Trees enable the efficient verification of large data volumes. A single root hash can validate the presence of any number of data elements.
2. Sensitivity to Data Modification: Owing to the properties of hash functions, altering even a single bit of the original data leads to a significant change in the root hash, facilitating the detection of any data modifications.
3. Selective Disclosure with Merkle Proofs: These allow for the verification of specific data elements without the need to disclose the entire dataset.

These advantages underpin the widespread use of this technology in various applications, including modern blockchain projects. However, it is also crucial to acknowledge the limitations of this technology, stemming from the inherent drawbacks of the Merkle Tree and Merkle Proof concept:
1. Computational Resources: While the construction and verification of Merkle Trees are relatively efficient, they still require significant computational resources, especially for large data volumes. For instance, if the size of $d_i$ is several hundred mega(giga)bytes, the initial step of proving $d_i \in D$ involves computing the hash code $H(d_i)$, requiring substantial computations. These calculations are necessary every time the proof $d_i \in D$ needs to be invoked.
2. Complexity in Merkle Patricia Trees: In this variant, the authentication path formation is more complex than in the classic Merkle Tree - hashes are augmented with prefixes generated using the RLP serialization algorithm. These prefixes are necessary for correctly ordering the hashes, ensuring that even if the leaves are reordered, the tree root remains unchanged. However, this mechanism requires additional computations and increases the volume of auxiliary data (for storing and processing the prefixes).

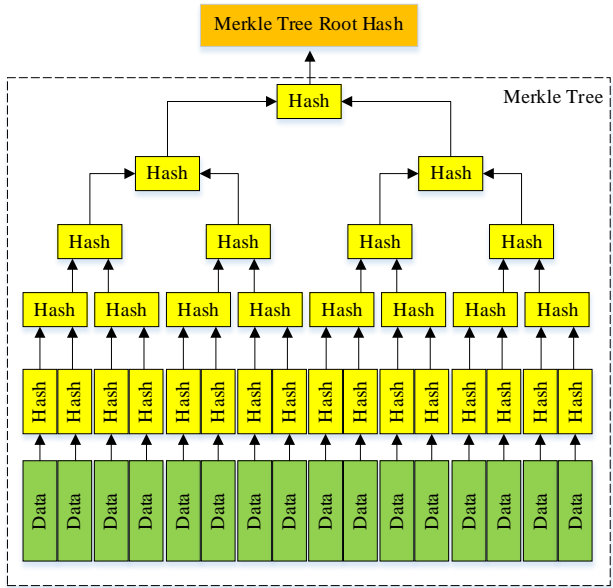

**FIGURE 1.** Merkle tree generation scheme

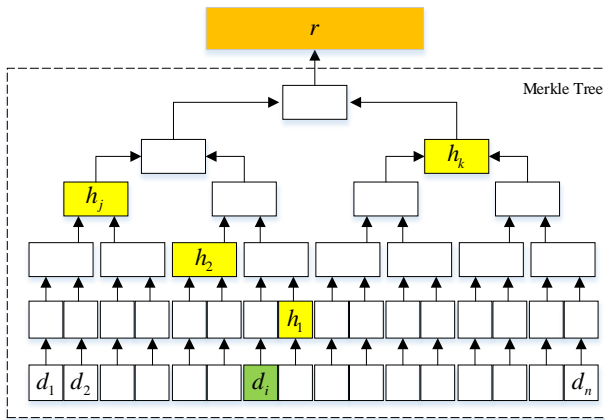

**FIGURE 2.** Example of using Merkle Proof

3.  Limited Functionality: Merkle Trees only allow for the verification of the presence of data elements and do not provide more complex functions, such as checking properties or conditions of the data.

The primary drawback of the native solution through Merkle Proofs is the length of the proof and the high cost of its implementation in a public blockchain. The Merkle Proof for each leaf in the tree is a path $h = \{h_1, h_2, ..., h_k\}$ (the authentication path or Merkle path). The path consists of all intermediate hashes required to generate the root from a particular leaf. The number of such hashes is determined by the number of tree levels, for example, 16 hashes for a tree with $2^{16}$ leaves. In the case of a 256-bit hash function (like Keccak-256), a 4096-bit (512-byte) path is required for each leaf in the tree. However, for larger Merkle Trees (containing $2^{56}$ or more leaves), the path length will be 14336 bits or more (at least 1792 bytes). The cost of publicly publishing such a proof as a smart contract would be prohibitively high.

In Ethereum, according to its Yellow Paper [35], [36], all data are stored in slots, and storing one 256-bit data slot requires 20,000 gas units. Considering that 8 bits equal one byte, one slot corresponds to 32 bytes (256 bits / 8 = 32 bytes). Therefore, 1 kilobyte (KB), which equals 1024 bytes, contains 32 slots (1024 bytes / 32 bytes = 32 slots). Consequently, storing 1 KB of data requires 640,000 gas units (32 slots * 20,000 gas per slot). Additionally, including the file in the transaction data field incurs extra gas costs. Since each byte of data costs 16 gas units, the total gas for 32 slots is 16,384 units (32 slots * 32 bytes * 16 gas per byte). Including the base transaction fee of 21,000 gas units, the total gas needed to store 1 KB of data is 677,384 units (640,000 + 16,384 + 21,000). At a gas price of 0.00000005 ETH per unit, storing 1 KB of data costs approximately 0.0338692 ETH (677,384 * 0.00000005). With the current ETH price around $2000, the cost of storing 1 KB of data on Ethereum is roughly $67.74.

Thus, the cost of storing a single path for a tree with $2^{16}$ leaves would be about $30, and for a tree with $2^{56}$ leaves, it would exceed $100. Even if these data are not stored on the blockchain but only used as metadata, the cost of such a solution remains high. For example, according to Ethereum's Yellow Paper, the charge for each byte is 68 gas, meaning the cost for one path for a tree with $2^{16}$ leaves would be about $3.2, and for a tree with $2^{56}$ leaves, more than $10. The situation worsens when it's necessary to prove the integrity of several data sets $d_i, d_j, ..., d_m$ included in one or several Merkle Trees. In this case, the data volume and computation costs multiply.

To overcome these drawbacks, this document proposes a new concept using ZKP systems (zk-SNARKs or zk-STARKs). The general idea involves applying ZKP technology to demonstrate that there is a valid Merkle Proof for a specific data element. In this case, we obtain a concise and quickly verifiable proof of the integrity of data $d_i$, based on the existence of a valid Merkle Proof with root $r$. Essentially, our goal is to reduce the size of the native Merkle

Proof and simplify the verification process, as the verifier does not need to store and submit all hashes $h = \{h_1, h_2, ..., h_k\}$ in the Merkle Proof. Instead, they can quickly verify a concise zk-proof by submitting only the first $H(d_i)$ and root $r = H(H(H(H(d_i) \| h_1)...) \| h_k)$ hashes. The proposed solution could also enhance privacy, as the verifier does not need to disclose all the data to verify the proof.

### C. ZERO-KNOWLEDGE PROOFS

ZKPs are a cryptographic method that allows one party (the prover) to convince another party (the verifier) of the truth of a statement without revealing any information other than the fact of the statement's truthfulness.

The main stages of generating and verifying ZKPs include [34]:

1.  Statement Formulation: The prover formulates the statement they wish to prove without revealing its content.
2.  Commitment: The prover sends a "commitment" - an encrypted message that will later be used to confirm the truthfulness of the statement.
3.  Challenge: The verifier sends a randomly generated request or "challenge" to the prover. This request determines the evidence to be provided.
4.  Response: The prover generates a response using their secret data and the received challenge. This response should prove the truth of the statement without revealing the statement itself or any additional information.
5.  Verification: The verifier analyzes the response and determines whether it corresponds to the challenge and the statement. If the response matches, the statement is considered proven.

In mathematical terms, ZKP can be represented as follows [34]:

*   Let $P$ be the prover, and $V$ be the verifier.
*   $P$ wishes to prove knowledge of a secret $x$, such that $y = f(x)$, where $f$ is a known function.
*   $P$ sends $V$ a proof $\pi$ that they know $x$, without revealing $x$.

This process ensures that $V$ learns only that $P$ knows $x$, but does not gain any information about $x$ itself. This is the essence of ZKP - proving knowledge without revealing the knowledge itself.

In modern theory and practice, ZKPs utilize both interactive and non-interactive proofs [34]:

*   Interactive ZKPs involve a series of exchanges between the prover and verifier. These exchanges consist of a sequence of challenges and responses, allowing the verifier to be convinced of the prover's statement's truthfulness without gaining any additional information. Interactive proofs are often considered more secure as they involve dynamic information exchange and can be adapted for various types of statements and scenarios. However, they require the presence of both parties in the proof process, and repeated data exchanges can be time-

consuming and require significant computational resources.

- Non-interactive ZKPs allow the prover to create a one-time proof, which can then be verified by any interested party without further communication with the prover. This approach does not require continuous communication between parties. Generated proofs can be verified by anyone at any time after their creation, making them particularly useful in blockchain technologies and other distributed systems. However, developing a reliable non-interactive proof can be more complex, and some types of statements may be impossible to prove in a non-interactive manner.

In our research, we utilize non-interactive ZKPs, focusing on blockchain applications where scalability, efficiency, and the widespread dissemination of proofs are crucial. This is also particularly relevant in the context of ensuring a high level of confidentiality and security.

### D. GENERATION OF ZERO-KNOWLEDGE PROOFS

To expand and conceptualize the basic stages of generating and verifying ZKPs, we will use terms and definitions from Dan Boneh's work [34].

The first step in generating ZKPs involves constructing a public arithmetic circuit, represented as a mathematical structure $C(x, w) \rightarrow F$, where $x$ is the public statement in the space $F^n$, and $w$ is the secret witness in the space $F^m$. In other words, the circuit $C$ computes a certain function based on $x$ and $w$, outputting the result in the field $C(x, w) \rightarrow F$.

The next step is the preprocessing (setup) of public parameters $(S_p, S_v)$, which will be used in the proof and verification process:

- $S_p$ are the public parameters for the Prover.
- $S_v$ are the public parameters for the Verifier.

The actual process of generating the proof

$$P(S_p, x, w) \rightarrow \pi$$

involves creating the proof $\pi$ using the public parameters $(S_p, S_v)$, the public statement $x$, and the secret witness $w$. Thus, the proof $\pi$ confirms the truth of the statement $x$ without revealing the secret witness $w$.

The verification process $V(S_v, x, \pi) \rightarrow \{accept\ or\ reject\}$ includes evaluating the proof $\pi$ by the verifier using their public parameters $S_v$ and the public statement $x$. As a result, the verifier either accepts the proof (confirming the truth of the statement) or rejects it (if the proof does not match the statement).

These stages represent the key components in the creation and verification of ZKPs, providing the ability to prove the truth of a statement without disclosing confidential information, a fundamental principle in cryptography and blockchain technology.

It is noteworthy that in the development of ZKP technology, the concepts of SNARKs (Succinct Non-Interactive Arguments of Knowledge) and STARKs (Scalable Transparent Arguments of Knowledge) have gained the most traction, differing primarily in the preliminary setup phase:

- SNARKs require a so-called "setup ceremony" or "trusted setup." During this setup, cryptographic keys are generated, which are then used to create and verify proofs. If these keys are compromised, the security of the entire system could be at risk.
- In contrast to SNARKs, STARKs do not require a specific setup ceremony and do not need a trusted setup. This makes them "transparent," meaning there is no need to generate and store secret keys, eliminating risks associated with their compromise.

Thus, the absence of a need for a trusted setup makes STARKs more resistant to certain types of attacks and simplifies their deployment, as there is no need to worry about the security of setup keys. However, SNARKs can offer shorter proofs and faster verification, which can be particularly useful in terms of deploying these technologies in blockchain systems.

### E. AGGREGATION OF ZERO-KNOWLEDGE PROOFS

Proof aggregation allows the combination of multiple proofs into a single one, simplifying the verification process. This leads to more efficient operations since the verifier does not need to check each proof individually [34].

Suppose we have $n$ different proofs $\{\pi_1, \pi_2, ..., \pi_n\}$ and their corresponding public inputs $\{x_1, x_2, ..., x_n\}$. In this case, each proof $\pi_i$ is generated using the function $P$ and corresponding parameters $(S_p, x_i, w_i)$, where $w_i$ are the private (or "secret") inputs.

The aggregated proof $\pi_{agg}$ can be represented as:

$$\pi_{agg} = P_{agg}(\{(S_p, x_i, w_i)\}, i = 1...n).$$

Here, $P_{agg}$ is the proof aggregation function. This function takes as input the sets of input data for each individual proof and generates an aggregated proof.

The public inputs in the aggregated proof are all the public inputs $\{x_1, x_2, ..., x_n\}$ of each of the original proofs. The private inputs in the aggregated proof are all the private inputs $\{w_1, w_2, ..., w_n\}$ of each of the original proofs.

During the verification of the aggregated proof, the function $V_{agg}$ is used, which takes as input the aggregated proof, public inputs, and verifier parameters $S_v$:

$$V_{agg}(S_v, \{x_i\}, i = 1..n, \pi_{agg}) \rightarrow \{accept\ or\ reject\}.$$

This approach to proof aggregation is pivotal in enhancing the scalability and efficiency of verifying multiple ZKPs. By aggregating proofs, we significantly reduce the computational burden on the verifier, especially in scenarios involving large datasets or numerous transactions. This aggregation methodology is particularly beneficial in blockchain systems, where reducing the computational and storage overhead can lead to more streamlined and cost-effective operations. Moreover, it aligns with the overarching goal of maintaining high security and privacy standards while optimizing the performance of blockchain networks.

### F. RECURSIVE ZK-SNARK PROOFS

Recursive zk-SNARK proofs involve using one zk-SNARK to verify another, creating a chain of proofs where each successive proof validates the previous one. This technique effectively consolidates multiple proofs into one, enhancing verification efficiency by negating the need to check each proof separately [34].

Consider having two proofs, $\pi_1$ and $\pi_2$, with corresponding public inputs $x_1$ and $x_2$, and private inputs $w_1$ and $w_2$. A recursive proof $\pi_r$ can be constructed, proving the knowledge of both $\pi_1$ and $\pi_2$:

$$\pi_r = P(S_p, (x_1, x_2), (\pi_1, \pi_2)).$$

Verification of $\pi_r$ involves checking it against the public inputs $x_1$ and $x_2$:

$$V(S_v, (x_1, x_2), \pi_r) \rightarrow \{accept \, or \, reject\}.$$

If $\pi_r$ is accepted, it implies the validity of both $\pi_1$ and $\pi_2$. This not only streamlines the verification process but also enhances the system's scalability and efficiency, making it particularly suitable for blockchain applications where computational resources are at a premium.

By employing the studied concepts of Merkle Trees, ZKPs, their aggregation, and recursive proofs, we are developing a system that effectively proves the inclusion of data in specific blockchain blocks, like Ethereum. Our approach focuses on reducing the volume of data required for verification, increasing processing speed, and enhancing system security. These methods significantly diminish the need for storing and processing large volumes of data, which is critically important for blockchain platforms with limited resources. As a result, the proposed system represents a powerful tool for improving verification and audit processes in blockchain while ensuring a high level of data privacy and protection.

## V. PROPOSED ZERO-KNOWLEDGE PROOF SCHEME

In the proposed ZKP scheme, we focus on the element-by-element aggregation of a chain of primary zk-proofs that validate the computational integrity of cryptographic hashing used in constructing a Merkle Tree. Conceptually, this scheme can be described as follows, using notations from previous sections:

1. For each leaf $d_i$ in $D = \{d_1, d_2, ..., d_n\}$, we have (or obtain from a trusted source, such as a full node in the Ethereum network) a Merkle Proof - a set of hashes $h = \{h_1, h_2, ..., h_j, ..., h_k\}$, such that starting from $h_i = H(d_i)$ (the hash of $d_i$), we can derive the root hash $r$ of the tree as per Equation (1). In the case of Merkle Patricia Trees, each $h_j$ in $h$ is understood as a serialized (prefixed) hash code, i.e., the ready result of the RLP algorithm, which is fed into the hashing function $H$ in the correct order to unambiguously compute the root hash $r$.

2. For each $h_j \in h$ in (1), we generate a proof of computational integrity using zk-SNARK/STARK systems:

$$\pi_j = P(S_p, (x_j, w_j)),$$

where

$$x_j = H(H(H(H(d_i) \| h_1) ...) \| h_j)$$

and

$$w_j = H(H(H(d_i) \| h_1) ...) \| h_j$$

for all $j = 1, 2, ..., k$.

Additionally, we also generate a proof

$$\pi_0 = P(S_p, (x_0, w_0)),$$

where $x_j = H(d_i)$ and $w_0 = d_i$.

This process can be schematically represented as shown in Figure 3.

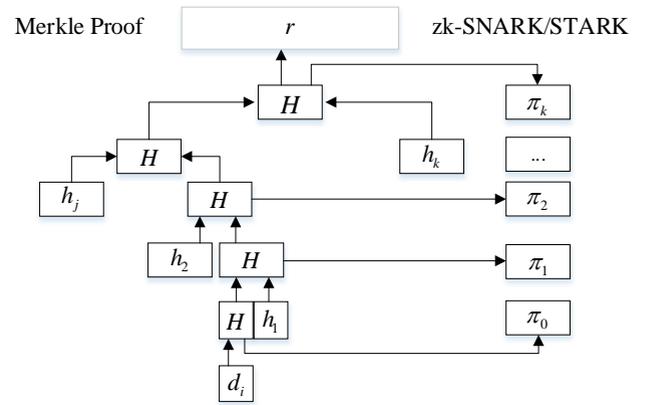

**FIGURE 3.** Schematic of computing the chain of proofs of computational integrity following the logic of the Merkle Tree

3. For all $j = 1, 2, ..., k$, we generate an aggregated (recursive) proof:

$$\pi r_j = P(S_p, (xr_j, wr_j)),$$

where

$$xr_j = H(H(H(H(d_i) \| h_1) ...) \| h_{j-1}), H(H(H(H(d_i) \| h_1) ...) \| h_j)$$

and

$$wr_j = \pi_j, \pi_{j-1}$$

for all $j = 1, 2, ..., k$.

The final step of recursion gives the proof

$$\pi r_k = P(S_p, (xr_k, wr_k)),$$

where

$$xr_k = H(H(H(H(d_i) \| h_1) ...) \| h_{k-1}), r,$$
$$r = H(H(H(H(d_i) \| h_1) ...) \| h_k),$$

and

$$wr_k = \pi_k, \pi_{k-1}$$

i.e., $\pi r_k$ is a proof of computational integrity for computing the root $r$ of the Merkle Tree. This process can be schematically represented as shown in Figure 4.

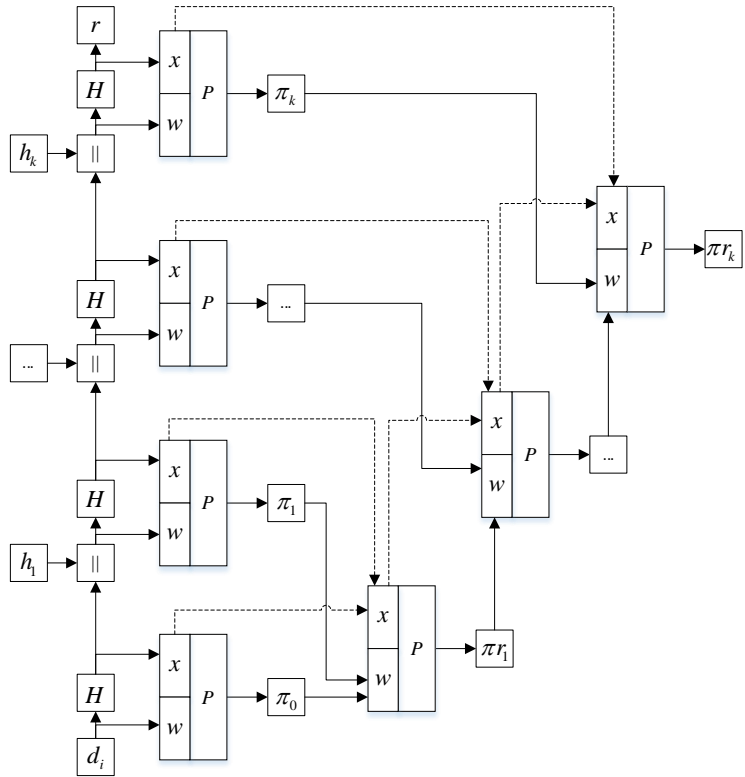

**FIGURE 4.** Schematic of generating the chain of proofs of computational integrity $\pi_0$, $\pi_1$, ..., $\pi_k$ and the chain of recursive proofs $\pi r_1$, $\pi r_2$, ..., $\pi r_k$ in Merkle Tree

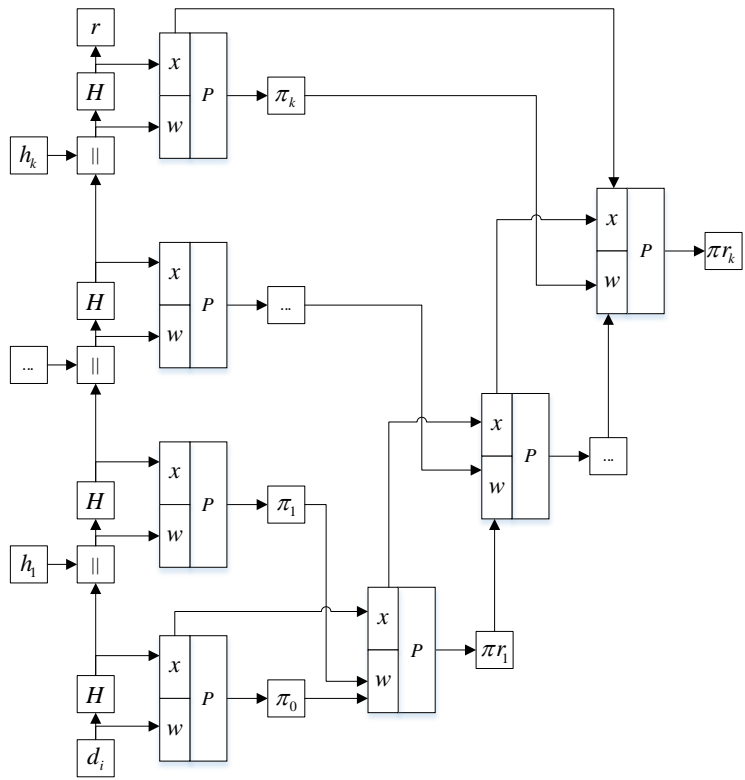

**FIGURE 5.** Simplified Scheme for Generating a Chain of Computational Integrity Proofs $\pi_0$, $\pi_1$, ..., $\pi_k$ and the Chain of Recursive Proofs $\pi r_1$, $\pi r_2$, ..., $\pi r_k$ in Merkle Tree

In our proposed ZKP scheme, the utilization of public inputs

$$xr_j = H(H(H(H(d_i) \| h_1)...) \| h_{j-1}), H(H(H(H(d_i) \| h_1)...) \| h_j)$$

in generating recursive proofs is optional (as indicated by italicized lines in the diagram). Providing all public inputs (along all italicized lines) would necessitate all intermediate hashes from the Merkle Proof $h = \{h_1, h_2, ..., h_j, ..., h_k\}$, concatenated with the result of the previous hashing. This approach, while comprehensive, is evidently redundant, and the scheme in Figure 4 can be employed in a simplified variant.

For instance, to prove the inclusion of leaf $d_i$ in $D = \{d_1, d_2, ..., d_n\}$, it is sufficient to submit only the first $H(d_i)$ and root $r = H(H(H(H(d_i) \| h_1)...) \| h_k)$ hashes as public inputs. The impossibility of falsifying intermediate proofs is ensured by the non-reversibility (resistance to pre-image) of the hashing function (refer to the following section). This simplified scheme is depicted in Figure 5.

Thus, we propose a novel approach that combines the advantages of Merkle Trees and zk-SNARKs/STARKs to ensure efficient and secure verification of data inclusion in a block, for instance, in the Ethereum blockchain. The core idea is to use zk-SNARKs/STARKs to create proofs of computational integrity for each step of hash computation in the Merkle Tree and then aggregate these proofs into a single recursive proof.

We note the following advantages of the proposed scheme:

- By aggregating proofs, we significantly reduce the amount of data and computation required for verification, enhancing the system's overall efficiency.
- The use of zk-SNARKs/STARKs ensures high levels of security and privacy, as the underlying details of the data and the proofs are not disclosed.
- The proposed method is highly scalable, suitable for large datasets and numerous transactions typical in blockchain systems.
- The simplified scheme in Figure 5 streamlines the verification process, making it more accessible and less resource-intensive.

In summary, our approach aims to harness the strengths of both Merkle Trees and advanced cryptographic proofs to offer a robust solution for data verification in blockchain environments, prioritizing efficiency, security, and scalability.

## VI. EXPERIMENTAL EVALUATION OF THE PROPOSED ZK-PROOF SYSTEM

This section presents the results of experimental testing of the developed zk-proof system integrated with Merkle Trees in the context of Ethereum blockchain. The experiment aims to assess the efficiency of the proposed scheme in terms of proof size, generation time, and verification time. The study is based on comparing the results of the scheme using zk-STARK and zk-SNARK technologies with traditional Merkle proof methods. The outcomes of this experiment provide key indicators that can be used to evaluate the system's suitability for practical application in modern blockchain networks.

### A. RESEARCH METHODOLOGY

Experimental investigations were conducted on a computational platform featuring an AMD EPYC 7003 processor, operating at speeds between 2.3 to 3.6 GHz, and equipped with 128 GB DDR4 2400 RAM. Our research focused on several critical metrics for the generation and verification of the final zk-proof $\pi r_k$, including:

- Generation Time of the Proof (seconds): This metric evaluates the computational efficiency of our zk-proof generation process.
- Verification Time of the Proof (seconds): This parameter assesses the feasibility and swiftness of verifying the generated proofs in real-world applications.
- Size of the Proof (bytes): A crucial aspect, especially significant in blockchain technologies where data storage and transmission efficiency are paramount.

We considered various scenarios corresponding to different lengths of the Merkle path $k$, with the native proof size for a binary Merkle tree being $32 \cdot k$ bytes (using a 32-byte hash function). However, the Ethereum blockchain employs Patricia-Merkle trees, where the size of the Merkle path is determined by the specific branch configuration. The size of the data in a particular leaf also depends on the specific block of transactions.

Our studies examined two types of trees: state trees and receipt trees. The effectiveness of our scheme was evaluated using several zk-proof technologies:

- First phase of the research focused on experiments with zk-STARK: This part of the study was dedicated to implementing the main proof scheme (see Fig. 5) for generating an aggregated proof. The objective was to measure the efficiency of this construction. The zk-STARK technology is renowned for its scalability and transparency, making it an attractive choice for extensive blockchain applications. However, a notable drawback is the larger proof size compared to zk-SNARK, which can impact its utility in systems with stringent data size constraints.
- The second phase of the research was devoted to further development using zk-SNARK technology: Specifically, we employed the Plonk and Groth16 protocols. Our goal at this stage was to achieve maximal compression of the final proof $\pi r_k$. The Groth16 protocol, in particular, is celebrated for generating short proofs, making it an ideal choice for systems where proof size and verification speed are critical. While zk-SNARK provides shorter proofs and faster verification than zk-STARK, it requires a trusted setup, which could be seen as a potential security risk.

Both phases of research significantly contributed to our understanding and optimization of zk-proof systems, particularly in the context of blockchain technologies like

Ethereum. The insights gained from these experiments are invaluable for advancing the field of cryptographic proofs and their practical applications.

### B. ZK-STARK EXPERIMENT RESULTS

The focus of this subsection is to present an analysis of the experimental results obtained from testing our zk-STARK-based proof system. This involved investigating two distinct types of Merkle trees - the receipt tree and the state tree - as they are typically employed in Ethereum blocks. Our objective was to examine the efficacy of zk-STARK in generating and verifying proofs, with an emphasis on the dynamics of proof size and computational efficiency.

For the experiments, we utilized a combination of real data from Ethereum blocks and synthetically generated data. The latter was particularly important for simulating scenarios with varying Merkle path lengths, allowing us to assess the scalability of our proof system.

The synthetic data for the state tree experiments followed a structured rule:

RLP(Keccak256(address), RLP(ethereum_account)),

where ethereum_account consists of:

- nonce: A randomly generated 256-bit value, simulating the number of transactions sent from an address.
- balance: A randomly generated 256-bit value, representing the account balance.
- storage_root: A simulated SHA3 hash of the StorageRoot, created by generating random 32-byte values.
- codeHash: A randomly generated 32-byte value, simulating the code hash associated with an Ethereum account.

This approach aimed to mimic real-world data structures in Ethereum, providing a more accurate assessment of our zk-STARK system's performance in a blockchain environment.

The outcomes of our experiments are detailed in Tables I, II, and III.

TABLE I
RECEIPT TREE EXPERIMENTS

| $k$ | Leaf Size (bytes) | Native Proof Size (bytes) | Generation Time (s) | Verificatio n Time (s) | Proof Size (bytes) | Native Verification Time (s) |
|---|---|---|---|---|---|---|
| 3 | 2071 | 783 | 5.8967 | 0.0911 | 152,996 | 9.2753e-6 |
| 5 | 963 | 516 | 5.7850 | 0.0905 | 152,996 | 6.2448e-6 |
| 6 | 3749 | 1371 | 5.9375 | 0.0933 | 152,996 | 1.616e-5 |
| 7 | 1435 | 1612 | 5.9304 | 0.0921 | 152,996 | 1.0763e-5 |

TABLE II
STATE TREE EXPERIMENTS

| $k$ | Leaf Size (bytes) | Native Proof Size (bytes) | Generation Time (s) | Verificatio n Time (s) | Proof Size (bytes) | Native Verification Time (s) |
|---|---|---|---|---|---|---|
| 5 | 147 | 2128 | 5.9493 | 0.0937 | 152,996 | 7.6899e-6 |
| 6 | 117 | 2179 | 5.9392 | 0.0931 | 152,996 | 8.1067e-6 |
| 7 | 112 | 2326 | 5.9581 | 0.0933 | 152,996 | 9.0437e-6 |
| 8 | 104 | 3596 | 6.1485 | 0.0940 | 152,996 | 1.1995e-5 |
| 9 | 113 | 3518 | 6.1029 | 0.0945 | 152,996 | 1.249e-5 |
| 10 | 116 | 3698 | 6.1502 | 0.0949 | 152,996 | 1.3464e-5 |

TABLE III
SYNTHETIC STATE TREE EXPERIMENTS

| $k$ | Leaf Size (bytes) | Native Proof Size (bytes) | Generation Time (s) | Verificatio n Time (s) | Proof Size (bytes) | Native Verification Time (s) |
|---|---|---|---|---|---|---|
| 4 | 139 | 1096 | 5.7772 | 0.0940 | 152,996 | 5.3183e-6 |
| 8 | 139 | 1829 | 5.8816 | 0.0930 | 152,996 | 8.2737e-6 |
| 12 | 139 | 2286 | 6.3720 | 0.0947 | 152,996 | 1.164e-5 |
| 16 | 139 | 4399 | 6.3696 | 0.0960 | 152,996 | 1.868e-5 |
| 20 | 139 | 5226 | 6.8266 | 0.0989 | 152,996 | 2.231e-5 |
| 24 | 139 | 5544 | 6.8185 | 0.0997 | 152,996 | 2.468e-5 |
| 28 | 139 | 7807 | 7.0227 | 0.1039 | 152,996 | 3.399e-5 |
| 32 | 139 | 8449 | 7.0049 | 0.1033 | 152,996 | 3.620e-5 |
| 36 | 139 | 10146 | 7.6580 | 0.1044 | 152,996 | 4.349e-5 |
| 40 | 139 | 10038 | 7.5672 | 0.1043 | 152,996 | 4.352e-5 |
| 44 | 139 | 11467 | 7.7270 | 0.1053 | 152,996 | 4.980e-5 |
| 48 | 139 | 13430 | 7.7724 | 0.1064 | 152,996 | 5.663e-5 |
| 52 | 139 | 13966 | 7.7834 | 0.1064 | 152,996 | 6.048e-5 |
| 56 | 139 | 14486 | 7.8025 | 0.1067 | 152,996 | 6.386e-5 |
| 60 | 139 | 16029 | 7.8501 | 0.1063 | 152,996 | 6.825e-5 |
| 64 | 139 | 17424 | 7.8410 | 0.1067 | 152,996 | 7.300e-5 |

The results demonstrate the practicality of zk-STARK for generating proofs of varying complexity. A key observation is the consistency of proof size across different path lengths, highlighting the scalability of our approach.

- For the receipt trees (Table I), we noticed a stable proof generation time, which suggests that our system can handle real-world Ethereum data efficiently.
- The state trees (Table II) presented a more varied scenario, with a slight increase in generation and verification times as the path length increased. This aligns with the expected computational complexity associated with larger Merkle paths.

The synthetic data experiments (Table III) were crucial in testing our system's performance under hypothetical, yet plausible, blockchain scenarios. These tests confirmed that our zk-STARK system maintains its efficiency and accuracy even as the complexity of the Merkle tree increases.

In conclusion, the zk-STARK experiments underscore the robustness and adaptability of our proof system in a blockchain environment. The consistent proof sizes, coupled with reasonable computational demands, make our approach a viable solution for real-world blockchain applications, particularly in the Ethereum ecosystem.

The zk-STARK experiment results, particularly focusing on the Ethereum blockchain blocks, highlighted a significant aspect of proof size in relation to the Merkle path length (k):

- In the Ethereum blockchain, the native proof size for a Merkle tree is determined by the path length (k) and the data structure used. For a binary Merkle tree, the native proof size is $k \cdot 32$, assuming a 256-bit hashing function. In the case of a Patricia-Merkle tree, used in Ethereum, the proof size is calculated based on the number of child nodes in the path. Each path element is $16 \cdot 32$ bytes plus 4 bytes of overhead data.
- Our experiments employed real block data from the Ethereum blockchain, ensuring that the results accurately reflect practical scenarios in one of the most prominent types of blockchain.

The experiments revealed that despite efficient generation and verification times, the zk-STARK proof sizes were larger than their native Merkle proof counterparts. This observation was consistent across various path lengths and in both receipt and state tree scenarios.

This realization of larger zk-STARK proof sizes compared to native proofs necessitated exploring zk-SNARK technology. The aim was to find a more efficient zk-proof system that could offer comparable or smaller proof sizes without compromising the system's efficiency and security. This approach was crucial for enhancing the practical application of our zk-proof system in blockchain technologies like Ethereum.

### C. ZK-SNARK EXPERIMENT RESULTS

In the subsequent phase of our research, we focused on the implementation and testing of zk-SNARKs, specifically Plonk and Groth16 protocols. The primary goal was to achieve the smallest possible proof size for efficient and cost-effective verification in the Ethereum network.

Using real Ethereum blockchain data, we generated zk-SNARK proofs for various scenarios within the Ethereum receipt and state trees. The key performance metrics evaluated were:

- Proof Generation Time: Time taken to generate the zk-SNARK proof.
- Proof Verification Time: Time required to verify the generated proof.
- Proof Size: The total size of the generated proof, a critical factor for storage and metadata use in Ethereum.
- Economic Analysis: Assessment of the cost implications of using zk-SNARKs for proof storage and metadata in Ethereum.

The results are summarized in Tables IV and V, illustrating the effectiveness of zk-SNARKs in reducing proof sizes and related costs.

Table IV outlines the results of zk-SNARK testing using real Ethereum receipt tree data. It focuses on the correlation between the path length (k) in the Merkle tree, the size of the native proof, and the performance metrics of the zk-SNARK proofs. Table V represents the testing results for the Ethereum state tree, illustrating how the zk-SNARK approach scales with various path lengths in a more complex Merkle tree structure. Analysis of the results shows:

- Groth16 consistently produced the smallest proofs (256 bytes), significantly smaller than the native proofs and even Plonk's proofs (928 bytes).
- Groth16's proofs are not only smaller but also more cost-effective, both for storage and metadata usage in Ethereum, compared to native proofs and Plonk.
- While Plonk provides considerable size reduction compared to native proofs, Groth16 offers even greater efficiency, making it an ideal choice for applications where proof size and cost are critical.

- With the current Ethereum prices, the cost savings using zk-SNARKs are substantial. This aspect is particularly important for applications involving frequent and large-scale transactions.

TABLE IV
ETHEREUM RECEIPT TREE ZK-SNARK RESULTS

| $k$ | Native Proof | | | Plonk's proofs (928 bytes) | | Groth16's proofs (256 bytes) | |
|---|---|---|---|---|---|---|---|
| | Size (bytes) | Storage Cost (ETH, $) | Metadata Usage Cost (ETH, $) | Storage Cost (ETH, $) | Metadata Usage Cost (ETH, $) | Storage Cost (ETH, $) | Metadata Usage Cost (ETH, $) |
| 3 | 783 | $51,80 | $5,32 | $61.39 | $6,31 | $16.94 | $1,74 |
| 5 | 516 | $34,13 | $3,51 | $61.39 | $6,31 | $16.94 | $1,74 |
| 6 | 1371 | $90,69 | $9,32 | $61.39 | $6,31 | $16.94 | $1,74 |
| 7 | 1612 | $106,64 | $10,96 | $61.39 | $6,31 | $16.94 | $1,74 |

TABLE V
ETHEREUM STATE TREE ZK-SNARK RESULTS

| $k$ | Native Proof | | | Plonk's proofs (928 bytes) | | Groth16's proofs (256 bytes) | |
|---|---|---|---|---|---|---|---|
| | Size (bytes) | Storage Cost (ETH, $) | Metadata Usage Cost (ETH, $) | Storage Cost (ETH, $) | Metadata Usage Cost (ETH, $) | Storage Cost (ETH, $) | Metadata Usage Cost (ETH, $) |
| 5 | 2128 | $140,77 | $14,46 | $61.39 | $6,31 | $16.94 | $1,74 |
| 6 | 2179 | $144,15 | $14,81 | $61.39 | $6,31 | $16.94 | $1,74 |
| 7 | 2326 | $153,87 | $15,81 | $61.39 | $6,31 | $16.94 | $1,74 |
| 8 | 3596 | $237,88 | $24,44 | $61.39 | $6,31 | $16.94 | $1,74 |
| 9 | 3518 | $232,72 | $23,91 | $61.39 | $6,31 | $16.94 | $1,74 |
| 10 | 3698 | $244,63 | $25,13 | $61.39 | $6,31 | $16.94 | $1,74 |

This comparative analysis underscores the economic viability of employing zk-SNARK proofs, particularly Groth16, in Ethereum. The drastically reduced proof sizes not only offer computational efficiency but also translate into substantial cost savings, both in terms of storage and metadata usage. The implications of these findings are profound, especially considering the ongoing scalability and cost-efficiency challenges in blockchain technologies. The blend of Plonk's flexibility and Groth16's succinctness presents a compelling case for their wider adoption in various blockchain applications.

## VII. SECURITY ANALYSIS OF THE PROPOSED ZK-PROOFS AGGREGATION SCHEME

This section is dedicated to evaluating the security dynamics of our novel aggregation scheme for zk-proofs within a Merkle Tree structure. The inherent replication of the binary Merkle Tree's architecture in our scheme prompts a security consideration based on the probability of root collisions. Specifically, we scrutinize the likelihood of scenarios where a calculated Merkle root $r'$ coincides with the actual root $r$ for a given Merkle path $h = \{h_1, h_2, ..., h_k\}$. Here, we present both precise analytical expressions and approximate formulations to gauge this probability.

A pivotal aspect of our analysis involves contemplating a hypothetical scenario where an adversary has the capability to manipulate not only the data $d_i \neq d_j$ but also the root $r$ itself, and potentially even the path elements $h = \{h_1, h_2, ..., h_k\}$. This exploration is crucial for comprehending how our system safeguards data and thwarts potential threats, especially in the context of Ethereum blockchain application.

By examining these factors, we aim not only to affirm the robustness and reliability of the proposed scheme but also to identify potential avenues for its further refinement and advancement. This scrutiny is integral to ensuring the system's resilience against evolving cryptographic challenges and maintaining its efficacy in blockchain-based applications.

### A. CRYPTOGRAPHIC PROPERTIES OF HASH FUNCTIONS

Cryptographic hash functions, denoted as $H$, are a fundamental element of many contemporary cryptographic systems. They transform input data $d_i$ of arbitrary length into fixed-length output data $h_i$ and must meet a series of stringent cryptographic requirements:

1. Collision Resistance. This property is defined by the impossibility of efficiently finding two different input values $d_i$ and $d_j$ such that $H(d_i) = H(d_j)$. Mathematically, for all $d_i \neq d_j$, the probability $P(H(d_i) = H(d_j))$ should approach zero.
2. One-Way Functionality. A hash function must ensure one-way secrecy, meaning that from the hash $h_i$, it should be computationally infeasible to efficiently determine the input $d_i$. Formally, for each hash $h_i$ and a randomly chosen $d_i$, it should be impossible to find a $d_j$ such that $H(d_j) = h_i$, except by brute force.
3. Pre-Image Resistance. This property requires that it should be computationally infeasible to find any input $d_j$ that hashes to a given hash $h_i$, meaning finding $d_j$ such that $H(d_j) = h_i$ should be unachievable in a reasonable time frame.
4. Second Pre-Image Resistance. This property ensures that for a given input $d_i$ and its hash $h_i = H(d_i)$, finding another input $d_j$, not equal to $d_i$ but satisfying $H(d_j) = h_i$, should be computationally infeasible. This requirement reinforces protection against attacks aimed at creating two different documents with identical hash values. Mathematically, it can be expressed as: for all $d_i$, finding $d_j \neq d_i$ such that $H(d_i) = H(d_j)$ should be computationally unfeasible.

This last property complements collision resistance, ensuring that even with the original input value $d_i$ and its hash, searching for an alternative input $d_j$ leading to the same hash remains an unattainable task. This reinforces the confidence in the uniqueness and non-replicability of each hash, which is critically important for maintaining the integrity and security of data in blockchain systems and applications requiring authentication of information.

These properties are achieved through complex algorithmic constructions that employ cryptographic compression methods. A key element is the fact that even minimal changes in input data lead to significant and unpredictable changes in the hash. This can be represented as:

$H(d_i + \Delta d) \neq H(d_i) + \Delta h$ where $\Delta d_i$ is a small change in the input data, and $\Delta h$ is the change in the hash.

Thus, hash functions provide a high level of uncertainty, excluding the possibility of reverse engineering or predicting the input data. By combining these properties, cryptographic hash functions provide a reliable foundation for data protection and information authentication, serving as a key tool in building blockchain networks and ZKP systems.

### B. PROBABILISTIC CHARACTERISTICS OF IDEAL HASH FUNCTIONS

When analyzing cryptographic hash functions that generate an $m$-bit hash, it is essential to start with the assumption of an ideal hash function. In this ideal model, every possible hash is assumed to be equally probable, and the hash function is presumed to distribute outputs uniformly across the entire hash space. Under this assumption, we examine the following probabilistic characteristics:

1. Collision Probability: With an $m$-bit output hash function, the total number of possible unique hashes is $2^m$. Applying the Dirichlet principle (or "birthday paradox"), the probability that two different inputs will yield the same hash, or the collision probability, can be approximated as:

$$P_{\text{collision}} \approx 1 - e^{-s(s-1) \cdot 2^{-(m+1)}}. \qquad (2)$$

Here, $s$ represents the number of hashing attempts. This equation demonstrates that even with a relatively low number of attempts, the likelihood of a collision becomes significant, highlighting the criticality of an adequate hash size. The derivation of formula (2) is given in Appendix A.

2. Pre-Image Resistance Probability: Searching for a pre-image entails finding an input $.d_j$. that hashes to a predetermined hash $h_i$. For an $m$-bit hash, the probability of successfully identifying a pre-image on a single trial is:

$$P_{pre-image} = 2^{-m}. \qquad (3)$$

This is derived from the assumption that the hash function distributes hashes evenly.

3. Second Pre-Image Resistance Probability: The second pre-image search involves identifying an alternate input $d_j$, distinct from $d_i$, so that $H(d_i) = H(d_j)$. The probability of this occurrence is analogous to the pre-image probability and equals:

$$P_{second\ pre-image} = P_{pre-image} = 2^{-m}. \qquad (4)$$

Under the assumption of an ideal hash function, these probabilistic attributes underscore the significance of selecting hash functions with sufficiently expansive hash sizes for cryptographic reliability. The collision probability, amplified by the birthday paradox, accentuates the need for hash functions with extensive hash spaces. Therefore, to maintain system security, employing hash functions that offer minimal collision likelihood and robust resistance to both pre-image and second pre-image searches is essential. This ensures data integrity and defends against data manipulation, vital for systems that leverage blockchain technologies and ZKPs.

### C. PROBABILISTIC ANALYSIS OF MERKLE TREES

In considering a Merkle Tree, let's assume the use of an ideal cryptographic hash function, denoted as $H$, which generates an $m$-bit hash code $H(d_i)$ for input data $d_i$. By 'ideal', we imply that the function generates all possible outputs with an equal probability of $2^{-m}$. Thus, the probability that two distinct inputs $d_i \neq d_j$ produce the same hash output $H(d_i) = H(d_j)$ is $2^{-m}$.

Consider a Merkle Tree with the root $r$. For arbitrary input data $d_i$ and its associated Merkle path $h = \{h_1, h_2, ..., h_k\}$, the root $r$ is computed by the rule:

$$r = r(d_i) = H(H(...H(H(H(d_i) \| h_1) \| ...) \| h_{k-1}) \| h_k).$$

Let's estimate the probability $P(r = r')$ that for two different inputs $d_j \neq d_i$, their corresponding roots coincide $r = r'$, where

$$r' = r(d_j) = H(H(...H(H(H(d_j) \| h_1) \| ...) \| h_{k-1}) \| h_k).$$

For $k = 1$, we have:

$$r = r(d_i) = H(H(d_i) \| h_1),$$
$$r' = r(d_j) = H(H(d_j) \| h_1).$$

The equality $r = r'$ is satisfied:
- when $H(d_i) = H(d_j)$,
- or (when $H(d_i) \neq H(d_j)$) when $H(H(d_i) \| h_1) = H(H(d_j) \| h_1)$.

Given that

$$P(H(d_i) = H(d_j)) \big|_{d_j \neq d_i} =$$
$$= P(H(H(d_i) \| h_1) = H(H(d_j) \| h_1)) \big|_{H(d_i) \neq H(d_j)} = 2^{-m},$$

we derive:

$$P(r = r') = 2^{-m} + (1 - 2^{-m}) 2^{-m}.$$

For $k = 2$, we consider:

$$r = r(d_i) = H(H(H(d_i) \| h_1) \| h_2),$$
$$r' = r(d_j) = H(H(H(d_j) \| h_1) \| h_2).$$

The equality $r = r'$ holds true under the following conditions:
- when $H(d_i) = H(d_j)$,
- or (when $H(d_i) \neq H(d_j)$) when $H(H(d_i) \| h_1) = H(H(d_j) \| h_1)$,
- or (when $H(d_i) \neq H(d_j)$ and $H(H(d_i) \| h_1) \neq H(H(d_j) \| h_1)$) when $H(H(H(d_i) \| h_1) \| h_2) = H(H(H(d_j) \| h_1) \| h_2)$,

thus leading to:

$$P(r = r') = 2^{-m} + (1 - 2^{-m}) 2^{-m} + (1 - 2^{-m})^2 2^{-m}.$$

Generalizing for an arbitrary path length $k$, we conclude:

$$P(r = r') = 2^{-m} + 2^{-m} \sum_{i=1}^{k} (1 - 2^{-m})^i.$$

In advancing the probabilistic analysis of Merkle Trees, we delve into the summation $\sum_{i=1}^{k} (1 - 2^{-m})^i$ as the sum of the first $k$ terms of a geometric progression. This sum can be represented as:

$$\sum_{i=1}^{k} (1 - 2^{-m})^i =$$
$$= 1 - 2^{-m} \frac{1 - (1 - 2^{-m})^k}{1 - (1 - 2^{-m})} =$$
$$= (2^m - 1)(1 - (1 - 2^{-m})^k).$$

Substituting into the formula for $P(r = r')$, we arrive at:

$$P(r = r') = 2^{-m} + (1 - 2^{-m})(1 - (1 - 2^{-m})^k). \quad (5)$$

Graphical representation, as shown in Figure 6, illustrates the dependency in equation (5) for $k, m = 1, .., 16$. From this three-dimensional graph, it's evident that for small $m$, the probability in equation (5) escalates swiftly. However, for larger $m$ and smaller $k$, the second term in equation (5) is close to zero, implying that $P(r = r')$ is only slightly greater than the pre-image and second pre-image probabilities outlined in equations (3) and (4).

Using the approximation $e^{-x} \approx 1 - x$ for small $x$, we reformulate $P(r = r')$ as:

$$P(r = r') \approx 2^{-m} + e^{-2^{-m}}(1 - (e^{-2^{-m}})^k).$$

For larger values of $m$, we consider $2^{-m} \approx 0$ and $e^{-2^{-m}} \approx 1$, leading to:

$$P(r = r') \approx 1 - e^{-k 2^{-m}}. \quad (6)$$

Figure 7 visualizes equation (6) for $k = 1, .., 64$ and $m \in \{16, 32, 48\}$. Dotted lines in Figure 7 also indicate $2^{-m}$ values as the probabilities of finding pre-images as per equations (3) and (4). This visualization reveals that even for significantly large Merkle Trees with $k = 64$ (having $2^{64}$ leaves), the increase in collision probability (5) compared to (3) and (4) is practically negligible.

As we delve deeper into the probabilistic characteristics of Merkle Trees, let's explore the graphical representations and implications of collision probabilities $P(r = r')$ based on different parameter settings:
- Figure 6 visualizes the dependence of $P(r = r')$ on parameters $k, m = 1, .., 16$.
- Figure 7 displays the probability variation of $P(r = r')$ concerning parameters $k = 1, .., 64$ and $m \in \{16, 32, 48\}$.

Considering $m = 256$, the following shifts in probability $P(r = r')$ are observed:
- For $k = 0$: $P(r = r') \approx 0.9 \cdot 10^{-77}$;
- For $k = 16$: $P(r = r') \approx 1.5 \cdot 10^{-76}$;
- For $k = 32$: $P(r = r') \approx 2.8 \cdot 10^{-76}$;
- For $k = 48$: $P(r = r') \approx 4.2 \cdot 10^{-76}$;
- For $k = 64$: $P(r = r') \approx 5.6 \cdot 10^{-76}$.

These calculations demonstrate that the probability of collision within Merkle Trees remains critically low across practically significant scenarios. For example, in Ethereum's blockchain, the Merkle path rarely exceeds $k = 56$, indicating that the likelihood of substituting original data $d_j$ with false data $d_i \neq d_j$ remains below $10^{-75}$.

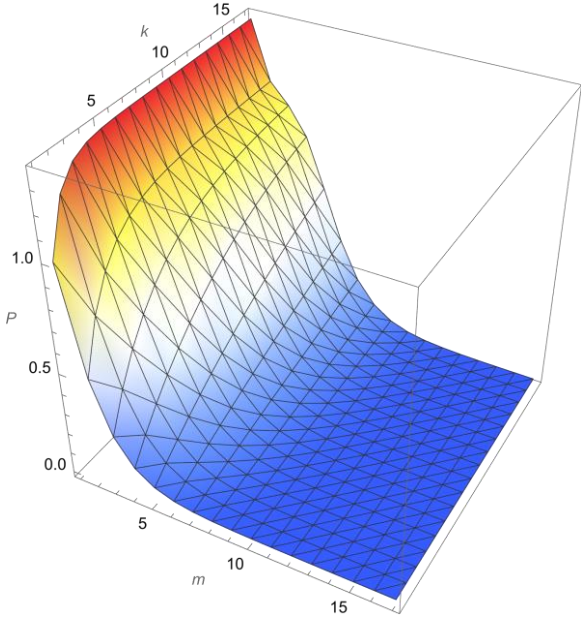

**FIGURE 6.** Dependency of Collision Probability $P(r = r')$ on Parameters $k$ and $m$ Ranging from 1 to 16

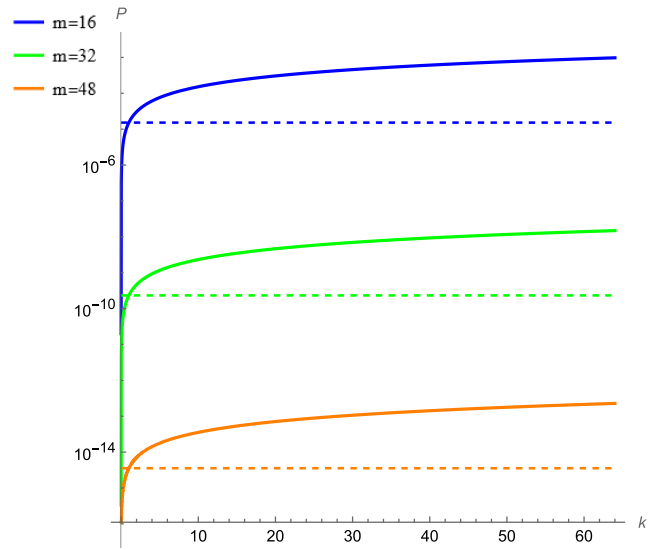

**FIGURE 7.** Variation of Collision Probability $P(r = r')$ Across $k = 1$ to 64 and $m$ Values of 16, 32, and 48

The above estimations assume a fixed root $r$ and path $h$. In a hypothetical scenario where an attacker can modify not only the data $d_i \neq d_j$ but also the root $r$ and potentially the path elements $h$, the formulas (5) and (6) need to be adjusted with $2^{-m/2}$ replacing $2^{-m}$. This aligns with the birthday paradox scenario, aiming to find any data-hash pairs that result

in a collision. Under this adjusted scenario for $m = 256$, we observe the following probability changes:

- For $k = 0$: $P(r = r') \approx 2.9 \cdot 10^{-39}$;
- For $k = 16$: $P(r = r') \approx 0.5 \cdot 10^{-37}$;
- For $k = 32$: $P(r = r') \approx 0.9 \cdot 10^{-37}$;
- For $k = 48$: $P(r = r') \approx 1.4 \cdot 10^{-37}$;
- For $k = 64$: $P(r = r') \approx 1.9 \cdot 10^{-37}$.

Thus, the proposed zk-proof system, leveraging the concept of Merkle Trees, ensures a high level of security. The probabilities of substituting original data with false data remain critically low, posing no significant threat to the integrity and security of the system. These findings affirm the robustness of the zk-proof system in maintaining data integrity, even under scenarios that extend beyond the typical use case in blockchain technologies like Ethereum. The versatility and resilience of this system underscore its potential in safeguarding data in diverse cryptographic applications.

## VIII. DISCUSSION OF RESEARCH FINDINGS

1. Proposed zk-Proof Scheme. Our analysis suggests that the proposed zk-proof scheme, integrating Merkle Trees with zk-SNARK/STARK technologies, represents a significant advancement in cryptographic protection and efficiency. The system not only ensures a high degree of data security and confidentiality but also offers improved scalability and versatility in handling various data structures. Particularly noteworthy is the scheme's ability to efficiently process large volumes of data, which is crucial for a wide range of applications, especially in the context of blockchain technologies.

2. Size and Verification Speed of zk-Proofs. Experimental results indicate that the sizes of the final zk-proofs are significantly reduced compared to traditional approaches. This reduction not only lessens storage and data transmission requirements but also enables more efficient network resource usage, particularly in the Ethereum blockchain. The rapid proof generation and verification times underscore the system's potential for real-time and high-performance applications.

3. Compatibility with Ethereum Structures. Testing results for Ethereum's receipt and state trees demonstrate our scheme's high compatibility with the current blockchain infrastructure. This confirms the feasibility of integrating our system into Ethereum's existing architecture, offering more efficient and secure transaction and smart contract processing.

4. Security Analysis of the Scheme. Probability assessments of collisions and data substitution in Merkle Trees confirm the robustness of the proposed scheme against attacks. The low probabilities of collisions, even with substantial tree sizes, underscore the scheme's reliability in practical usage scenarios. Furthermore, analyses of hypothetical scenarios where an attacker could adjust root and intermediate hashes confirm the scheme's resilience against more complex attacks. These results provide additional evidence of the system's real-world robustness and underline its significance for data security in blockchain applications.

Overall, the research findings validate that the proposed zk-proof scheme is a potent tool for ensuring security and efficiency in blockchain technologies. By combining the advantages of Merkle Trees, zk-SNARK, and zk-STARK, it opens new avenues for optimizing blockchains and expanding their applications across various domains.

## IX. CONCLUSION

In this paper, we have introduced a newly developed zk-proof scheme that integrates the concepts of Merkle trees with zk-SNARK/STARK technologies. This scheme significantly enhances security and efficiency in data processing, particularly in the context of blockchain technologies like Ethereum.

A key achievement of this scheme is the substantial reduction in the size of zk-proofs compared to traditional methods. This not only saves storage and data transmission resources but also accelerates the proof verification process, which is crucial for real-time systems and high-performance applications.

Experimental results confirm the high compatibility of our proposed scheme with the existing Ethereum architecture, opening possibilities for its integration into the current blockchain infrastructure for more effective and secure transaction and smart contract processing.

Our security analysis, including collision probability assessment and data substitution in Merkle trees, affirms the reliability of the scheme for practical application. The low probabilities of collisions, even for large trees, underscore the scheme's robustness in real-world usage scenarios.

Overall, the outcomes of this research demonstrate that the proposed zk-proof scheme is a powerful tool for ensuring security and efficiency in blockchain technologies. By combining the advantages of Merkle trees, zk-SNARKs, and zk-STARKs, it paves the way for blockchain optimization and expands its applications across various domains.

## APPENDIX A
## COLLISION PROBABILITY

Consider a cryptographic hash function producing an $m$-bit hash. The total number of unique hash outputs is $2^m$. When examining the likelihood of collisions, we consider the probability that two random inputs will not result in a collision. For the first pair of hashes, this probability is $1 - 2^{-m}$. As we select more hashes, the probability of the next hash not causing a collision decreases. For instance, for the second pair, this probability is $1 - 2 \cdot 2^{-m}$, and so forth.

After selecting $s$ hashes, the probability that the next hash will not result in a collision becomes $1 - (s-1) \cdot 2^{-m}$. Consequently, the probability that there will be no collision after selecting $k$ hashes is the product of all these individual probabilities:

$$P_{\text{no collision}} = \prod_{i=0}^{s-1} \left(1 - i \cdot 2^{-m}\right).$$

The collision probability, then, is the complement of this probability:

$$P_{\text{collision}} = 1 - P_{\text{no collision}} = 1 - \prod_{i=0}^{s-1} \left(1 - i \cdot 2^{-m}\right).$$

Approximation Using Exponential Functions: The product used in equation $P_{\text{no collision}}$ can be approximated using the concept of natural logarithms. The natural logarithm base $e$ is defined by the limit:

$$\lim_{n \to \infty} \left(1 + \frac{1}{n}\right)^n = e.$$

For large $m$ values, the fraction $2^{-m}$ is quite small. Hence, we can utilize the approximation $e^{-x} \approx 1 - x$ for small $x$:

$$P_{\text{no collision}} \approx \prod_{i=0}^{s-1} e^{-i \cdot 2^{-m}}.$$

This approximation is valid because $e^{-x}$ is the inverse function of $\left(1 + \frac{1}{n}\right)^n$ as $n \to \infty$. The product of exponentials is equivalent to the exponential of a sum, so:

$$P_{\text{no collision}} \approx e^{-\sum_{i=0}^{s-1} i \cdot 2^{-m}}.$$

Using the formula for the sum of an arithmetic progression, which is $s(s-1)/2$, we have:

$$P_{\text{no collision}} \approx e^{-s(s-1) \cdot 2^{-(m+1)}}.$$

This derivation leads us directly to equation (2), providing a clear understanding of the collision probability in cryptographic hash functions. This mathematical framework is crucial for evaluating the security and integrity of hash-based cryptographic systems.

To establish the "birthday bound" in the context of probability theory and cryptography, let's dissect the problem into a series of logical steps to derive the necessary formulas. Our objective is to ascertain the number of samples $k$ required to achieve a given probability $p$ of collision in a system with $2^m$ possible outcomes.

Starting from expression (2), we aim to solve for $k$ given a collision probability $p$:

$$p \approx 1 - e^{-k(k-1) \cdot 2^{-(m+1)}}.$$

Rearranging and simplifying this equation, we get:

$$e^{-k(k-1) \cdot 2^{-(m+1)}} \approx 1 - p,$$

$$-k(k-1) \cdot 2^{-(m+1)} \approx \ln(1 - p),$$

$$k(k-1) \approx -2^{m+1} \ln(1 - p).$$

For large $k$, $k(k-1)$ can be approximated as $k^2$, thus:

$$k^2 \approx -2^{m+1} \ln(1 - p),$$

$$k \approx \sqrt{-2^{m+1} \ln(1 - p)}.$$

Particularly for $p = 0.5$, the equation simplifies to:

$$k \approx \sqrt{-2^{m+1} \ln(0.5)} = \sqrt{2^{m+1} \ln(2)} .$$

Given that $\ln(2)$ is approximately 0.693, the equation becomes:

$$k \approx 1.1774\sqrt{2^m} \approx 2^{m/2} .$$

This deduction implies that to achieve a computational complexity comparable to brute-force search, the hash length must be doubled. In other words, if an adversary is capable of computing $2^{m/2}$ hash values through brute-force, they will begin to encounter hash collisions for all hashes shorter than $m$ bits.

This calculation elucidates a critical aspect of cryptographic hash functions - the relationship between the size of the hash and its vulnerability to collisions. As $m$ increases, the effort required to find a collision increases exponentially, reinforcing the importance of using sufficiently large hash sizes in cryptographic applications.

Moreover, the "birthday bound" provides a theoretical benchmark for evaluating the security of hash functions against collision attacks. It demonstrates that the probability of finding a collision grows significantly with the square root of the number of possible hash outcomes. This insight is pivotal in cryptographic design, guiding the choice of hash lengths to ensure robust security against collision-based vulnerabilities.

## ACKNOWLEDGMENT

This project is supported by Proxima Labs, 1501 Larkin Street, suite 300, San Francisco, USA.

## DECLARATION OF INTERESTS

We declare that the authors have no competing financial interests, or other interests that might be perceived to influence the results and/or discussion reported in this paper. The results/data/figures in this manuscript have not been published elsewhere, nor are they under consideration (from you or one of your Contributing Authors) by another publisher. All of the material is owned by the authors and/or no permissions are required.

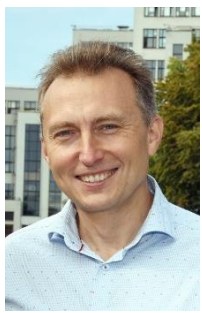

**OLEKSANDR KUZNETSOV** (Member, IEEE) holds a Doctor of Sciences degree in Engineering and is a Full Professor. He is an Academician at the Academy of Applied Radioelectronics Sciences and the recipient of the Boris Paton National Prize of Ukraine in 2021. Additionally, he serves as a Visiting Professor at the Department of Political Sciences, Communication and International Relations at the University of Macerata in Italy. He is also a Professor in the Department of Security Information Systems and Technologies at the V. N. Karazin Kharkiv National University, Ukraine. His research primarily focuses on applied cryptology and coding theory, blockchain technologies, the Internet of Things (IoT), and the application of AI in cybersecurity.

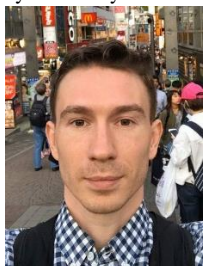

**ALEX RUSNAK** is a seasoned investor and the lead of technology projects at Proxima Labs, located at San Francisco, USA. With over 8 years of specialized expertise in blockchain technologies, Mr. Rusnak has established a significant presence in the field. His expertise extends to in-depth research in cryptography, a domain where he has authored several notable research papers. He possesses a profound understanding and experience in advanced cryptographic protocols, particularly Starky, Plonky2, and Plonky3. Mr. Rusnak's contributions in these areas have been instrumental in the advancement of cryptographic solutions and blockchain technology.

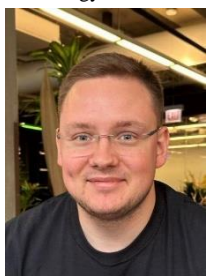

**ANTON YEZHOV**, leading the product and project management teams at Proxima Labs, has made substantial contributions to the field of privacy protocols, including the development of Penumbra. He played a pivotal role in implementing the Near zk light client and has co-authored numerous research papers in this domain. Prior to his foray into the crypto space, Mr. Yezhov had an illustrious career in artificial intelligence, where he was instrumental in developing the first AI voice systems. His diverse background in AI and cryptography underscores his comprehensive expertise in both fields.

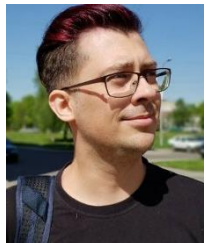

**DZIANIS KANONIK** is an esteemed Engineering Leader with over 12 years of experience in applied cryptography, mathematics, circuit design, electronics, big data, and AI. Before joining Proxima Labs, Mr. Kanonik led technical teams dedicated to the development of AI-driven high-frequency trading systems. He has also been involved in researching computer vision algorithms, contributing significantly to the advancement of self-driving technology. His extensive experience and leadership in these technically demanding fields have positioned him as a key figure in the industry.

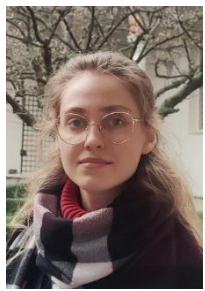

**KATERYNA KUZNETSOVA** is cryptographer with Master degree in Cybersecurity (2023, V. N. Karazin Kharkiv National University). She has been working in the field of cryptography and zero-knowledge proof technology. She is the author of several scientific papers in cryptography and cybersecurity. Currently, she serves as a leading research programmer at Proxima Labs. Kateryna's work focuses on the development and exploration of high-performance zero-knowledge proofs, a vital component in ensuring privacy and security in cryptographic systems. Her contributions in this area demonstrate her expertise and commitment to advancing cybersecurity measures.

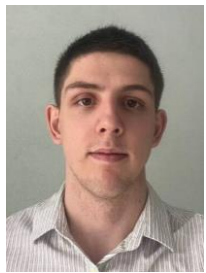

**STANISLAV KARASHCHUK** is an emerging talent in the field of cryptographic security, currently pursuing his Bachelor's degree in Mathematics and Computer Science at the Faculty of Mechanics and Mathematics, Taras Shevchenko National University of Kyiv. Since 2022, Mr. Karashchuk has been deeply involved in the realm of cryptographic security and zero-knowledge proof technology. His expertise in these areas is further amplified by his current role as a research programmer at Proxima Labs. Mr. Karashchuk's work primarily focuses on the development and research of high-performance zk-proofs. His dedication and innovative approach to problem-solving in the field of cryptographic security demonstrate his potential as a significant contributor to the advancement of cryptographic methods and technologies.